\documentclass[pra,12pt,tightenlines,showpacs,showkeys]{revtex4}

\usepackage{hhline}
\usepackage{graphicx}
\usepackage{bbm}
\usepackage{amsfonts}
\usepackage{theorem}
\usepackage{amssymb}
\usepackage{amsmath}

\theoremstyle{plain}
\newtheorem{thm}{THEOREM}

\newcommand{\version}{\today}

\newcommand{\beq}{\begin{equation}}
\newcommand{\eeq}{\end{equation}}
\def\beqa{\begin{eqnarray}}
\def\eeqa{\end{eqnarray}}

\newcommand{\C}{{\mathbb C}}
\newcommand{\W}{{\mathcal W}}
\newcommand{\third}{\mbox{$\frac{1}{\sqrt{3}}$}}
\newcommand{\R}{{\mathbb R}}
\newcommand{\one}{{\mathbbm 1}}
\newcommand{\Tr}{{\rm Tr}}

\newcommand{\G}{{\mathcal G}}

\date{\small\version}

\begin{document}

\title{\begin{flushright}\begin{small}UWthPh--2006--13\end{small}\end{flushright}
The state space for two qutrits has a phase space structure in its
core}
\author{Bernhard Baumgartner\footnote{Bernhard.Baumgartner@univie.ac.at},
 Beatrix C. Hiesmayr\footnote{Beatrix.Hiesmayr@univie.ac.at}, Heide Narnhofer\footnote{Heide.Narnhofer@univie.ac.at}
} \affiliation{Institut f\"ur Theoretische Physik, Universit\"at
Wien, Boltzmanngasse 5, 1090 Vienna, Austria}

\begin{abstract}
We investigate the state space of bipartite qutrits. For states
which are locally maximally mixed we obtain an analog of the
``magic'' tetrahedron for bipartite qubits---a magic simplex $\W$.
This is obtained via the Weyl group which is a kind of
``quantization'' of classical phase space. We analyze how this
simplex $\W$ is embedded in the whole state space of two qutrits and
discuss symmetries and equivalences inside the simplex $\W$. Because
we are explicitly able to construct \textit{optimal} entanglement
witnesses we obtain the border between separable and entangled
states. With our method we find also the total area of
\textit{bound} entangled states of the parameter subspace under
intervestigation. Our considerations can also be applied to higher
dimensions.


\end{abstract}

\maketitle


\section{Introduction}\label{intro}

In 1935 Erwin Schr\"odinger stated already that ``entanglement is
the quintessence of the quantum theory''. The late discoveries and
developments in many distinct branches of physics show its immense
validity. It is the basis for quantum cryptography, quantum
teleportation and maybe if realizable quantum computation. It has
also triggered a new field: quantum information.


The main problem for composite systems is how to find out if a given
state is separable or not and thus to characterize the border
between separability and entanglement. While we have for the
simplest composite system---two two--level systems ($2\times 2$
systems) or bipartite qubits--- a necessary and sufficient criterion
for separability, the Peres criterion, it is for higher dimensions
only a necessary criterion (except $2\times 3$). The criterion
states that every separable density matrix is mapped into a positive
semidefinite matrix by partial transposition (PT), i.e., by a
transposition on one of the subsystems. The reason why it fails for
higher dimensions is that for these systems a completely positive
map cannot be characterized by transposition alone and moreover
these systems show more aspects of entanglement.

It is obvious that the knowledge of the state space is the key
ingredients to understand entanglement and therefore for developing
and optimizing new applications. Moreover it will help in
understanding the relation of different entanglement measures.

In this paper we focus on bipartite qutrits ($3\times 3$ systems).
We concentrate on the set of locally mixed states with a
quasiclassical structure and construct a geometrical picture of the
state space. The quasiclassical structure fits also exactly into the
conditions needed for teleportation and dense coding, e.g.
Refs.~\cite{BW92,W01,W05}. While these sets of states have been
noted already in Ref.~\cite{VW00, N06}, only little is known about
its structure concerning entanglement, witnesses, PPT (positive
partial transposition) and possible bound entanglement.

For two qubits four orthogonal Bell states can be used to decompose
every locally mixed state and a geometric picture can be drawn. In
such a geometrical approach the Hilbert-Schmidt metric defines a
natural metric on the state space, e.g. Ref.~\cite{B05,Ovrum}. Via
diagonalizing every locally mixed state can then be described by
three real parameters which can be used to identify the density
matrix by a point in a $3$--dimensional real space. The positivity
condition forms a tetrahedron with the Bell states at the corners
and the totally mixed state, the trace state, in the origin. Via
reflection one obtains another tetrahedron with reflected Bell
states at the corners, see e.g. Ref.~\cite{BNT02}. The intersection
of both simpleces gives an octahedron where all points inside and at
the border represent separable states because they are the only ones
invariant under reflection and thus positive under PT. While for
qubits this characterizes the separable set of locally mixed states
fully we show in this paper that the analogue to the octahedron for
qutrits is not quite that simple and in addition not all locally
maximally mixed states can be imbedded.

The simplex for bipartite qutrits lives in a $9$--dimensional
Euclidean space where the borders are given by the positivity
condition of density matrices. We construct two polytopes and prove
that they are an inner (kernel polytope) and an outer fence
(enclosure polytope) to the border of separability. The boundary
achieved by taking the set of all states which are positive under PT
has not only linear faces and corners but also curved parts. Then we
explicitly show how to construct optimal witnesses and apply them to
certain states and show that there are regions where there is bound
entanglement, i.e. entanglement which cannot be distilled by local
operation and classical communication (LOCC).

The paper is organized as follows, we present first the construction
of the set of states we are analyzing, the magic simplex $\W$. Then
we discuss how the set is embedded in the whole set of states. We
proceed with analyzing the rich structure of symmetries inside $\W$:
The symmetry of a discrete classical phase space. Hereupon we focus
on describing the boundary of separability by calculating optimal
witnesses. The optimization is done analytically and also
numerically. Further we added an appendix for more details.

Many of our considerations can be extended to pairs of qudits. In order to be as concrete as possible we postpone
generalizations to higher dimensions and more abstract analyzes to a following companion paper.

\section{The construction of the magic simplex $\W$}\label{const}

Throughout this paper we focus on two parties with $3$ degrees of
freedom e.g. ``qutrits''. Take any maximally entangled pure state
vector in the Hilbert space $\C^3 \otimes \C^3$ for defining a
``Bell type state''. Denote this vector as $\Omega _{0,0}$. Choose
the bases $\{|0\rangle , |1\rangle , |2\rangle \}$ in each factor
such that \beq \label{omega} \Omega _{0,0} = \third \sum_s |s\rangle
\otimes |s\rangle\, . \eeq On the first factor in the tensorial
product -- the side of Alice -- we consider actions of the
\textbf{Weyl operators}. They are defined by \beqa \label{weyl}
W_{k,\ell}|s\rangle & = & w^{k(s-\ell)}|s-\ell\rangle ,   \\
w & = & e^{2\pi i/3}\, . \eeqa Throughout this paper the letters
$\{s,t,j,k,\ell,m,n,p,q\}$ denote the numbers $0, 1, 2$.
Calculations with them are to be understood as ``modulo 3''. So
\quad ``$1+2$''$=0$, \quad ``$2\times 2$''$= 1$, \quad ``$-1$''$=2$,
\quad etc.\quad

The transformations which we consider take place on the first
factor, on the side of Alice. Bob's side is in our definitions
inert. This is not really an asymmetry. For the Bell state
$\Omega_{0,0}$ every action of an operator $A$ on the side of Alice
is equivalent to the action of a certain $\tilde{A}$ on the side of
Bob. Concerning the Weyl operators the equivalent action is
$\tilde{W}_{k,\ell} = w^{-2k\ell} W_{k,-\ell}$. Changing the roles
of Alice and Bob i.e. the flip transformation is therefore
equivalent to a local reflection combined with a phase factor, but
with no change of the total set of the produced states. The phase
factors will disappear in the projection operators to be defined in
equation (\ref{proj}), and the reflection is one of the symmetries
studied in Sect.~\ref{inside}.

The actions of the Weyl operators -- we simplify the notation and write
$W_{k,\ell}|\Omega \rangle$ meaning $( W_{k,\ell}\otimes \one)|\Omega \rangle$ --
produce on the whole nine Bell type state vectors
\beq
\Omega_{k,\ell} = W_{k,\ell}\Omega_{0,0}\, .
\eeq
The Weyl operators obey the {\bf Weyl relations}
\beqa \label{Weylrel}
W_{j,\ell}W_{k,m} & = & w^{k\ell}W_{j+k,\ell +m} \, ,         \\
W_{k,\ell}^\dag  = W_{k,\ell}^{-1} & = & w^{k\ell}W_{-k,-\ell} \, ,       \\
W_{0,0} & = & \one \,. \eeqa We remark that the Weyl operators and
the unitary group which they form appear sometimes in disguise,
under names like ``generalized spin operators'', ``Pauli group'' and
``Heisenberg group'', Refs.~\cite{G98,PR04}.

The original use of the Weyl operators, in the chapter
``Quantenkinematik als Abelsche Drehungsgruppe'' of Ref.~\cite{W31}
was the ``quantization'' of classical kinematics. (Both continuous
and discrete groups have their appearance there.) In the appendix we
present a physics model for the bipartite system of qutrits, which
may help to visualize the ideas, concerning the interplay of
quasiclassical and quantum structures.

The set of index pairs ${(k,\ell )}$ is the {\bf discrete classical
phase space}; $\ell$ denotes the values for the coordinate in
``x-space'', $k$ the values of the ``momentum'', see also
Fig.~\ref{phasespace}.

To each point in this space is associated a projection operator \beq
\label{proj} P_{k,\ell} =
|\Omega_{k,\ell}\rangle\langle\Omega_{k,\ell}| \,. \eeq This
projection operator is the density matrix for a Bell type state. The
mixtures of these pure states form our object of interest, the {\bf
magic simplex} \beq \W\quad =\quad \{ \quad \sum
c_{k,\ell}P_{k,\ell} \quad |\quad c_{k,\ell}\geq 0 , \quad \sum
c_{k,\ell}=1 \quad \}\,, \eeq with the nine pure states $P_{k,\ell}$
at the corners. As a geometrical object it is located in an
8-dimensional hyperplane of the 9-dimensional Euclidean space $\{A=
\sum a_{k,\ell}P_{k,\ell} \quad |\quad a_{k,\ell}\in \R\}$, equipped
with a distance relation $\sqrt{\Tr (A-B)^2}$. Specifying the origin
$A=0$ in this Euclidean space, it is also equipped with a norm, the
Hilbert-Schmidt norm $\sqrt{A^2}$, and the inner product $\Tr (A B)
= \sum a_{k,\ell}\,b_{k,\ell}$.

The {\it geometric} symmetry of this simplex  for the qutrits is
larger than the symmetry which is related to the underlying
algebraic relations. The later one is equal to the symmetry of the
classical phase space. This is studied in Sect.~\ref{inside}.

\section{How is $\W$ embedded in the set of states?}\label{place}

$\W$ contains only states which are locally maximally mixed, i.e.
every partial trace gives the unit matrix times the normalizing
constant. Further it contains the maximal possible number of
mutually orthogonal pure states. While this characterization is
sufficient for qubits, defining the magical tetrahedron or any
locally unitary transform of it, this is not so for the qutrits.
More explicitly for qubits every locally maximally mixed state can
be embedded into a magical tetrahedron, while for the qutrits we
observe:
\begin{enumerate}
\item \label{undec}
There exist locally maximally mixed states, which cannot be
diagonalized with maximally entangled pure states, the Bell type
states.
\item \label{unw}
Even if such a maximally mixed state is decomposable into orthogonal Bell type states,
it may be inequivalent to any of the states in $\W$.
\item \label{ninebell}
There are maximal sets of nine mutually orthogonal Bell states,
which do not build an equivalent of $\W$.
\end{enumerate}
Examples are presented in the appendix.

We remark that there are other ways to characterize the density
matrices, by expanding them into products of operators which are a
basis for the space of matrices. The use of products of Weyl
operators in Ref.~\cite{N06} is closely related to the construction
in this paper. And it is the analogue to the use of products of
Pauli matrices, e.g. Ref.~\cite{BNT02}, considered as forming a
group. Another method has been tried, considering the analogue of
Pauli matrices as generators of $SU(2)$. This leads to using the
Gell-Mann matrices, e.g. Ref.~\cite{B05}, generators of $SU(3)$.

There are several ways how to characterize a special unitary equivalent of one of the
versions of $\W$. One is already given by the construction: Choose one of
the Bell states, and choose some basis on one side. Another way would be a choice of
fixed special Bell states which have to be represented. There is a precise statement about
the restrictions and the freedom to do this:
\begin{thm}\label{bellandw}
Every pair of mutually orthogonal Bell states can be embedded into a version of $\W$.
Such a pair fixes the appearance of a certain third Bell state.
A fourth state can then be embedded, if it is orthogonal to the first three.
Then, with four different Bell states, all elements of $\W$ are fixed.
\end{thm}
\textbf{PROOF} Choose one vector out of the pair as $\Omega_{0,0}$.
Take a Schmidt decompositions of this vector and of $\Phi$ the
second one with the same basis on Bob's side:
$$\Omega_{0,0}= \third \sum_s |\phi_s \rangle \otimes  |\eta_s \rangle,
\quad \Phi= \third \sum_s |\psi_s \rangle \otimes  |\eta_s \rangle .$$
Consider the unitary operator $U=\sum_s |\psi_s \rangle \langle  \phi_s |$,
acting on the first factor.
Orthogonality of the Bell states implies
$$ 3\,\langle \Omega_{0,0}\, |\, \Phi\rangle = \Tr\, U = 0\;.$$
This is possible only if the eigenvalues of U are the three numbers
$\{e^{i\delta}w^k \}$ with some common phase factor $\delta$. Now
let $\{|s\rangle\}$ be the eigenvectors of $U$, and fix
$W_{1,0}=e^{-i\delta}U$. This implies $P_{1,0}= |\Phi \rangle
\langle\Phi |$. Note, that there are still three phase factors not
fixed, one for each basis vector. Nevertheless, the projector
$P_{2,0}=W_{1,0}P_{1,0}W_{1,0}^\dag$ is defined unambiguously. The
fourth Bell state has a Schmidt decomposition $\third \sum_s |\chi_s
\rangle \otimes  |s \rangle$. Observe that the orthogonality to the
first three states implies $\sum_s w^{ks}\langle s|\chi_s \rangle
=0$ for each $k$, and so $\langle s|\chi_s \rangle =0$ for each $s$.
Together with the orthogonality of the $\chi_s$, this implies that
either $|\chi_s \rangle = e^{i\eta(s)}|s+1\rangle$, or $|\chi_s
\rangle = e^{i\eta(s)}|s+2\rangle$. Now fix the phases for each
$|s\rangle$, such that all $\eta (s)=\eta$, and the fourth Bell
state is implemented as either $P_{0,1}$ or as $P_{0,2}$. So all the
ingredients for the construction of $\W$ are fixed. $\Box$\\
\\
That the special choice of the positions in the phase space makes no
difference for the total set of elements is made clear by
consideration of the symmetries inside $\W$.

\section{Symmetries and equivalences inside $\W$}\label{inside}

We consider linear symmetry operations mapping $\W$ onto itself that can be
implemented by local transformations of the Hilbert space.
So separability remains unchanged.
We show that the transformations of $\W$ can be considered as transformations of
the quasi classical discrete phase space.

Letting the Weyl operators act on Alice's side gives the {\bf phase
space translations}: \beq\label{transl} \mathcal{T}_{m,n}: \quad
P_{k,l}\mapsto  P_{k+m,\ell+n} = W_{m,n}P_{k,\ell}W_{m,n}^\dag\,.
\eeq The action is a discrete ``Galilei transformation''. The
quantization, expressed in the phase factors of the Weyl relations,
disappears due to the combined action of $W_{m,n}$ and its adjoint.

In the appendix we present the Weyl operators in matrix form,
and also their relations to the phase space. As usual, we take the ``x-coordinate'' $\ell$
as {\it horizontal} and the ``momentum coordinate'' $k$ as {\it vertical}.

\begin{figure*}
\center{\includegraphics[width=250pt,keepaspectratio=true]{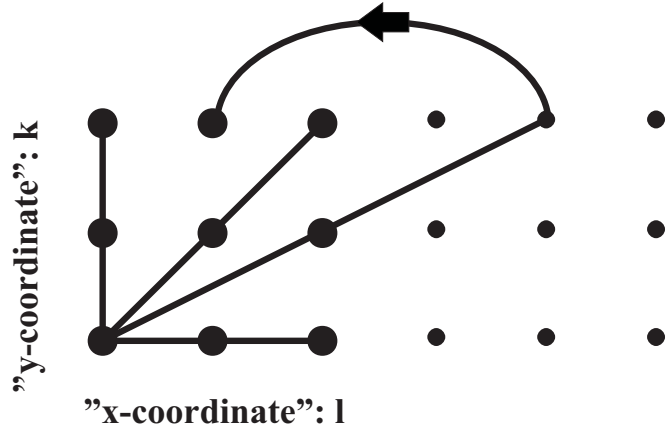}}
\caption{Here we plotted the points $P_{l,k}$ of the discrete
classical phase space. $l$ denotes the values of the $x$ coordinate
and runs from $0$ to $2$ and $k$ ``quantizes'' the momentum and runs
also from $0$ to $2$. From one fixed point, e.g. $P_{0,0}$, all
possible lines are drawn. Thus the phase space can be divided into
four bundles where each bundle consists of three parallel lines. In
Sect.~\ref{inside} we show that transformations inside the simplex
$\W$ are equivalent to transformations in this phase space and that
the lines are all equivalent in the sense that each line may be
transformed into any other one. This enables us to study the
geometry of separability in the magic simplex
$\W$.}\label{phasespace}
\end{figure*}

For all the other special operations $P_{0,0}$ stays fixed. This is
no general restriction, since translations may shift each of the
points in phase space to the origin. Now we have to use the help of
operators acting on Bob's side. For every linear operator $A$ on
Alice's side there exists an operator $\tilde{A}$ acting on the
other party, such that \beq \label{tilde} A|\Omega_{0,0}\rangle =
\tilde{A}|\Omega_{0,0}\rangle \,. \eeq They are related through
transposition in the preferred basis:
$$\langle s|A|t\rangle = \langle t|\tilde{A}|s\rangle .$$
So, for every local unitary $U$ \beq U  \tilde{U}^\dag
|\Omega_{0,0}\rangle = |\Omega_{0,0}\rangle \,. \eeq Thus every
unitary transformation of the Weyl operators can be lifted to a
unitary transformation of $\W$: \beqa
W_{m,n} &=& e^{i\eta}U W_{k,\ell} U^\dag ,\\
\Rightarrow \qquad  P_{m,n} &=& U \tilde{U}^\dag P_{k,\ell} U^\dag
\tilde{U} \,. \eeqa This follows in detail from \beqa\label{lift} U
\tilde{U}^\dag P_{k,\ell} U^\dag \tilde{U} &=& U \tilde{U}^\dag
W_{k,\ell} P_{0,0} W_{k,\ell}^\dag\tilde{U} U^\dag
= U W_{k,\ell} \tilde{U}^\dag P_{0,0}\tilde{U} W_{k,\ell}^\dag U^\dag \nonumber \\
&=& U W_{k,\ell}U^\dag U\tilde{U}^\dag P_{0,0} U^\dag \tilde{U} U
W_{k,\ell}^\dag U^\dag = W_{m,n} P_{0,0} W_{m,n}^\dag\,. \eeqa First
consider
$$U_R :  \quad |s\rangle \mapsto \third \sum_t w^{-st}\,|t\rangle$$
It effects the {\bf quarter rotation of phase space} (counter
clock-wise) \beq\label{rot} \mathcal{R}:  \quad P_{k,\ell}\mapsto
P_{\ell,-k}\,. \eeq Then consider $$U_V :  \quad |0\rangle \mapsto
|0\rangle ,\, |1\rangle \mapsto  |1\rangle ,\, |2\rangle \mapsto w^2
|2\rangle , $$ it lifts to the {\bf vertical shear}
\beq\label{vshear} \mathcal{V}:  \quad P_{k,\ell}\mapsto
P_{k+\ell,\ell}\,. \eeq Combined action perfects the {\bf horizontal
shear} \beq\label{hshear} \mathcal{H} =
\mathcal{R}^{-1}\mathcal{V}^{-1}\mathcal{R}:  \quad
P_{k,\ell}\mapsto  P_{k,\ell+k}\,. \eeq

Now, for the following reflections we have to consider {\it
anti-unitary} transformations of the Hilbert space - unless we want
to use the flip, the exchange of Alice's and Bob's side. The
simplest one in our preferred basis is the {\bf vertical reflection}
\beq\label{refl} \mathcal{S}:  \quad P_{k,\ell}\mapsto
P_{-k,\ell}\,. \eeq It is implemented by complex conjugations \beq
 \sum_s a_s |s\rangle \mapsto \sum_s a_s^{*} |s\rangle
\eeq on both factors. Their tensorial product gives global complex
conjugation \beq C:  \quad \sum_{s,t} a_{s,t} |s\rangle \otimes
|t\rangle \mapsto \sum_{s,t} a_{s,t}^{*} |s\rangle\otimes
|t\rangle\,. \eeq Obviously $$C^{-1}=C,\quad CW_{k,\ell}|\Psi\rangle
=W_{-k,\ell}C|\Psi\rangle,\quad P_{-k,\ell}=CP_{k,\ell}C.$$ We
notice that this anti-unitary transformation is also compatible with
the Weyl relations. Since it acts globally, on both factors, it is
positivity, separability and PPT preserving. All the structural
properties of $\W$ which are of interest are symmetric under
vertical reflection. Hence they are also symmetric under combined
action with other transformations, which give new kinds of
reflections, as {\bf horizontal reflection} \beq\label{hrefl}
\mathcal{R}\mathcal{S}\mathcal{R}^{-1}:  \quad P_{k,\ell}\mapsto
P_{k,-\ell} \eeq and {\bf diagonal reflection} \beq\label{rdia}
\mathcal{R}\mathcal{S}: \quad  P_{k,\ell}\mapsto P_{\ell,k}\,. \eeq

All the transformations are ``linear'' or ``affine'' mappings of the phase space.
{\bf Phase space lines} $((k,\ell), (k+m,\ell+m), (k+2m,\ell+2m))$ are mapped onto lines.
Note that the sequence of the three points in a line can be rearranged in any way.
With the right relabelling of the indices modulo three one gets again the special prescribed form.

Let us collect our results and state the following theorem:
\begin{thm}\label{linear}
The group of symmetry transformations of $\W$ is equal to the group
of affine transformations of the quasi classical phase space which
is formed by the indices, \beq \left( \begin{array}{c} k \\ \ell
\end{array} \right) \mapsto \left( \begin{array}{cc} m & n \\ p & q
\end{array} \right) \left( \begin{array}{c} k \\ \ell \end{array}
\right) + \left( \begin{array}{c} j \\ r \end{array} \right)\,, \eeq
where $mq-pn\neq 0$, with all calculations done with integers modulo
$3$. For $mq-pn=1$ the transformation of the Hilbert space is
unitary, for $mq-pn=-1$ it is anti-unitary.
\end{thm}
\textbf{PROOF} Pure phase space translation by $(j, r)$ is the
second part, combined with the unit matrix, $m=q=1,\quad n=p=0$. The
generating elements for the first part of transformations and the
corresponding matrices are (in this Proof we use ``$-1$'' for
``$2$'') \beqa
\textrm{Quarter rotation:} \qquad\mathcal{R} \quad &\leftrightarrow &\quad \left( \begin{array}{cc} 0 & 1 \\ -1 & 0 \end{array} \right),\\
\textrm{Vertical shear:} \qquad\mathcal{V} \quad &\leftrightarrow &\quad \left( \begin{array}{cc} 1 & 1 \\ 0 & 1 \end{array} \right),\\
\textrm{Vertical reflection:} \qquad\mathcal{S} \quad
&\leftrightarrow & \quad \left( \begin{array}{cc} -1 & 0 \\ 0 & 1
\end{array} \right)\,. \eeqa All the invertible matrices can be
generated. This can be seen first by looking at the numbers of
zeroes. Maximally two zeroes, in relative diagonal positions, are
possible, as in $\mathcal{R}$ and $\mathcal{S}$. One zero is
possible, as in $\mathcal{V}$. There may be different positions of
the zeroes, but they can be moved by the diagonal reflection,
applied from the left and/or from the right. The case with no zero
in the matrix is represented by $\mathcal{V}\mathcal{H}$. Finally
there are different distributions of signs, but they cannot be
changed individually, since this would not give invertible matrices.
The signs can be changed pairwise, for each column or row, by the
vertical reflection $\mathcal{S}$, and by
$\mathcal{R}\mathcal{S}\mathcal{R}^{-1}$, applied from the left
and/or from the right. $\Box$\\
\\
The group structure of the combined transformations can be written
in matrix notation: \beq \left(
\begin{array}{c}  k \\ \ell \\ 1 \end{array} \right) \mapsto \left(
\begin{array}{ccc}  m & n & j  \\ p & q &  r \\ 0&0&1 \end{array}
\right) \left( \begin{array}{c} k \\ \ell \\ 1 \end{array}
\right)\,. \eeq

The lines in the discrete phase space play also an important role in
connection with the Mutually Unbiased Bases, see
Refs.~\cite{PR04,W04,B04}. In Fig.~\ref{phasespace} we visualize all
possible lines for one phase space point. Thus we have for the whole
phase space four bundles---called ``striations'' or ``pencils'' in
Refs.~\cite{W04,B04}, respectively---, each one with three parallel
lines. This makes 12 special sets out of $\left(
\begin{array}{c} 9\\3\end{array}\right)=84$ subsets with three
points of the phase space. The lines are all {\it equivalent} in the
sense that each line may be transformed into any other one. More
general, we have
\begin{thm}\label{equi}
In the classes of subsets of phase space points, there is just one equivalence class
of single points, one of pairs, two classes of triples and two of quadruples.
The equivalence relations are moreover valid also for the complementary sets
with five to eight points.
Inside each pair and inside each triple, there is total symmetry under permutations.
\end{thm}
\textbf{PROOF} We move the subsets to special chosen places in phase
space. Consider one point of the subset after the other, in any
order. Translation brings the first one to the origin $(0,0)$. The
shear transformations bring the second one to $(1,0)$. If these two
are part of a line, the third one has then been moved automatically
due to the linearity of the transformations, together with the first
two, to the point $(2,0)$, completing this vertical line. If the
third (or fourth) point is not in a line with the first pair, it is
movable with horizontal reflection and vertical shear to the place
$(0,1)$, without changing the arrangements of the line with $\ell
=0$. In the case of four points, with the vertical line at $\ell=0$
not yet completed, we have several cases: If the fourth point is
either at $(2,2)$ or at $(0,2)$, it completes another line, and we
can start again, moving this line to the preferred vertical one, and
the remaining point as done above. In case the fourth point is not
yet at $(1,1)$, where we want to place it, completing a square, it
is either at $(2,1)$ or at $(1,2)$, and it can be moved by shear,
together with one of the others, to form the preferred square. These
are the cases, where no complete line is contained in the subset of
four.

The inner symmetries of pairs and triples are now implicitly proven,
since the sequence of moving their points can be chosen arbitrarily.
 $\Box$

\section{The geometry of separability}

We now proceed to the question of separability. We start with a
rough analyzes of an inner and outer fence in $\W$. Then we
concentrate on the border given by PPT. In particular we show that
the test for positivity under PT reduces to a check for positivity
of a $3\times 3$ matrix. As an example we study the entanglement of
mixtures of the total mixed state and two orthogonal Bell type
states. We then explicitly describe the construction of witnesses
and analyze the strategy to optimize them. We apply our method to
the above state and find for some mixtures bound entanglement. As an
further example we discuss a density matrix which is a mixture of
the total mixed state and three orthogonal Bell states where two of
them are equality weighted.

\subsection{Two polytopes as inner and outer fences for
separability}\label{polytopes}

The most mixed separable state, with density matrix $\omega =\frac19
\one$, lies at the center of $\W$. \beq \omega = \frac19
\sum_{k,\ell}P_{k,\ell}\,, \eeq since the $\Omega_{k,\ell}$ form an
orthonormal basis. The separable states with the largest distance to
the center are defined by the lines in the phase space.
\begin{thm} \label{twelve}
The twelve outermost separable states in $\W$ have the density
matrices \beq \rho_{line}= \frac13 \sum_{(k,\ell )\, \in \, line}
P_{k,\ell}\,. \eeq
\end{thm}
\textbf{PROOF} This is a special case of the more abstract general
statement in Ref.~\cite{N06} equation (36). For a more concrete
demonstration, consider the vertical phase space line $\{(k,\ell
)\}=\{(0,0),(1,0),(2,0)\}$: \beq P_{k,\ell=0}= \frac13 \sum_{s,t}
w^{k(s-t)}|s,s\rangle \langle t,t|\,, \eeq where we now write
$|s,t\rangle$ for $|s\rangle \otimes |t\rangle $. With $\sum_k
w^{k(s-t)}=3\delta_{s,t}$ one gets \beq \rho_{line}=\frac13 \sum_k
P_{k,0}= \frac13 \sum_s |s,s\rangle \langle s,s|\,, \eeq obviously a
separable state.

It lies in each one of the three hyperplanes $B_{p,0}$ in our
Euclidean space, defined by \beq B_{p,q}=\{(c_{k,\ell})|\, c_{p,q} =
\frac13 \} \label{isoplane1} \,.\eeq $B_{p,0}$ intersects the line
of isotropic states $(1-\alpha ) \omega + \alpha P_{p,0}$, exactly
at the border between separable and entangled states at $\alpha
=\frac14$, see also Ref.~\cite{VW00}. This hyperplane is therefore
the proper entanglement witness, reduced to our subspace of
hermitean matrices. Each state outside is entangled, and it is only
the center of the triangle with the $P_{p,0}$ at the corners which
is a separable state.

By the equivalence relations stated in Theorem~\ref{equi}, all these
considerations are valid for each one of the twelve phase space
lines. Now the witness hyperplanes intersect also at the centers of
the other 72 ($=84 - 12$) triangular faces, but the states there are
not separable. This is easily checked by showing
that they are not PPT. This is done explicitly in the next Sect.~\ref{timesnine}. $\Box$\\
\\
The nine pairs of hyperplanes
\beqa
B_{p,q}&=&\{(c_{k,\ell}) \; | \quad  c_{p,q} = \frac13\; \}  \label{isoplane2}\\
\textrm{and} \qquad A_{p,q} &=& \{(c_{k,\ell}) \; | \quad c_{p,q} =
0\; \} \eeqa enclose all the separable states of $\W$. They define
the \beq \textbf {enclosure polytope}\qquad \{ (c_{k,\ell})\;
|\quad\textrm{all}\;\; c_{p,q}\in [ 0,\frac13 ] \}\,. \eeq It has
the same geometric symmetry as the simplex $\W$, which has nine
corners. Intersections of the $B_{p,q}$-hyperplanes in triples give
84 vertices. In each of the $B_{p,q}$ there lie 28 of these
vertices, the other 56 lie in $A_{p,q}$.

Of these 84 vertices, only twelve are separable states.
All the convex combinations of these twelve are again separable states. They form the
\beq
\textbf {kernel polytope}\qquad \{ \rho = \sum_{lines \, \alpha} \lambda_\alpha \, \rho _{line \, \alpha}
 \quad |\quad
\lambda_\alpha \geq 0,\quad \sum \lambda_\alpha =1 \}\,. \eeq It has
twelve vertices, which are the $\rho_{line}$. Each hyperplane
$B_{p,q}$ contains four of them. The other eight are in $A_{p,q}$
and define a full seven-dimensional convex body. This may be
compared to the four triangular faces of the qubit-octahedron lying
inside the triangles of the magical tetrahedron. The other four, out
of all eight, lie in witness planes, see Refs.~\cite{HH96,BNT02}.
Here is one more of the many differences between qubits and qutrits
(compare with Ref.~\cite{VW99}): We do not have the geometric
rotation-reflection symmetry between the bordering planes. In the
witness-hyperplane $B_{p,q}$ there is only a three-dimensional face
(a tetrahedron) with four vertices of the kernel polytope.

That the bordering face in $A_{0,0}$ of the kernel polytope is seven-dimensional
can be seen by looking in the Euclidean space of
hermitean matrices at the eight vectors $v_{p,q}$, with $(p,q)\neq (0,0)$.
Such a vector $v_{p,q}$ is defined as pointing from
$\frac18 (\one - P_{0,0})$, the center of the face of $\W$ in $A_{0,0}$, to
$P_{p,q}$, one of the eight vertices of the seven-dimensional simplicial face of $\W$.
The center of this face of $\W$ is also the center of the eight vertices in $A_{0,0}$ of the kernel polytope.
The vector is now representable as a linear combination of the
$\rho_{line}$, phase space lines through $(p,q)$ not containing $(0,0)$.
By equivalences and symmetries it is sufficient to demonstrate this for one example:
\beqa
2v_{2,2} & = & - [(P_{1,0}+P_{2,1}+P_{0,2}) + (P_{0,1}+P_{1,2}+P_{2,0})] + \\
          && -\frac13[(P_{1,0}+P_{1,1}+P_{1,2}) + (P_{2,0}+P_{1,1}+P_{0,2}) + (P_{0,1}+P_{1,1}+P_{2,1})] + \nonumber \\
         && +\frac13[(P_{2,0}+P_{2,1}+P_{2,2}) + (P_{1,0}+P_{0,1}+P_{2,2}) + (P_{0,2}+P_{1,2}+P_{2,2})]\,. \nonumber
\eeqa
This consideration is, by equivalence, valid for the maximal face in any $A_{p,q}$.

Every set of phase space points characterizes a face of $\W$, with dimension one less than the number $N$ of points.
So the types of faces of the kernel polytope,
surfacing in a face of $\W$, correspond to equivalence classes of sets of phase space points.
For $N=7$ there are five vertices $\rho_{line}$, given by the different lines formed by
subsets of the $7$ prescribed points. It is not difficult to classify:
For $N=6$ there are two types, one type giving faces containing three vertices $\rho_{line}$, the other two.
For $N=5$ there are two types, one containing two vertices, the other only one.
For $N=4$ either one vertex or none is present; $N=3$ is either a phase space line,
giving one vertex at the symmetry center, or another triple with no kernel vertex.
Also the edges, $N=2$, contain no vertex.

Note that the facts about isotropic states of qutrits, and the
witness hyperplanes $B_{p,q}$ which we use, as above in
Theorem~\ref{twelve}, is found by us in a new way as a byproduct of
our special methods. See the next two Sect.~\ref{timesnine} and
Sect.~\ref{witness}, where we proceed to find out more about the
border between the separable and the entangled states.

\subsection{PT of our $9 \times 9$ matrices}\label{timesnine}

We represent the density matrices in the basis of product vectors
\beq |s-\ell , s\rangle = |s-\ell\rangle \otimes |s\rangle \eeq and
order them into groups of three, according to $\ell$. Inside each
group we order according to $s$. So the global Hilbert space is
represented as a direct sum \beq \C^3_{\ell = 0} \oplus \C^3_{\ell =
1} \oplus \C^3_{\ell = 2}\,. \eeq The projectors \beq
P_{k,\ell}=\frac13 \sum_{s,t} w^{k(s-t)} |s-\ell , s\rangle \langle
t-\ell , t| \eeq do not mix the subspaces. A general matrix of $\W$
splits therefore into the direct sum \beq \rho =
\sum_{k,\ell}c_{k,\ell} P_{k,\ell} = \left(  \sum_k c_{k,0} P_k
\right) \oplus
      \left(  \sum_k c_{k,1} P_k \right) \oplus  \left(  \sum_k c_{k,2} P_k \right)
\eeq with the $3 \times 3$ matrices \beq P_0 = \frac13 \left(
\begin{array}{ccc}  1 & 1 & 1  \\ 1 & 1 &  1 \\ 1&1&1 \end{array}
\right) , \quad P_1 = \frac13 \left( \begin{array}{ccc}  1 & w^{*} &
w  \\ w & 1 & w^{*} \\ w^{*}&w&1 \end{array} \right), \quad P_2 =
\frac13 \left( \begin{array}{ccc}  1 & w & w^{*} \\w^{*} & 1 & w \\
w&w^{*}&1 \end{array} \right)\,. \eeq

Partial transformation is now the linear mapping
\beqa
|s-\ell ,s\rangle \langle t-\ell , t|\quad  \mapsto \quad |t-\ell ,s\rangle \langle s-\ell , t|\quad
&=&\quad  |m-s,s\rangle \langle m-t, t| \\
\textrm{with} \qquad m &=& s+t-\ell \nonumber \,. \eeqa It produces
a new grouping of the basis vectors $|m-s,s\rangle$, according to
$m$, and a new splitting of the global Hilbert space as \beq \C^3_{m
= 0} \oplus \C^3_{m = 1} \oplus \C^3_{m = 2}\,. \eeq The most
general element of $\W$ is \beq \label{lsplit} \rho = A_{\ell =0}
\oplus A_{\ell =1} \oplus  A_{\ell =2} \eeq with \beq \label{defma}
A =\frac13 \left( \begin{array}{ccc}  d & a^{*} & a  \\ a & d &
a^{*}\\ a^{*}&a&d \end{array} \right) , \quad d_\ell = \sum_k
c_{k,\ell}, \quad  a_\ell = \sum_k w^k c_{k,\ell}\,. \eeq Partial
transposition maps, as is demonstrated in the appendix, such a
matrix $\rho$ into \beq \label{ptm} B\oplus B\oplus B,
\qquad B=\frac13 \left( \begin{array}{ccc}  d_0 & a_2 & a_1^{*} \\
                  a_2^{*} & d_1 & a_0\\ a_1 & a_0^{*} & d_2 \end{array} \right)
\eeq with three times the same $3\times3$ matrix. Thus a test for
PPT of $\rho$ reduces to a check for positivity of the matrix $B$.

Now we apply the method and use the Peres criterion Ref.~\cite{P96}:
PPT, the positivity under partial transposition, is a necessary
condition for separability. So NPT, non-positivity under partial
transposition, implies entanglement. The missing detail in the proof
of Theorem~\ref{twelve}, that the $72$ triangular faces of $\W$ not
corresponding to phase space lines contain no separable point, is
contained in the following
\begin{thm}\label{lowface}
Consider a five dimensional face $F$ of $\W$, opposite to a triangular face which
contains a separable point. $F$ is spanned by the six Bell type states $P_{k,\ell}$
which are not located on the phase space line giving the separable state in the
triangular face.
The only entangled states in $F$, including its bordering faces,
are two $\rho_{line}$, and the edge joining them.
\end{thm}
\textbf{PROOF} By equivalence, we may assume that it is the vertical
line with $\ell=2$, which gives the separable state in the
triangular face, and which stays empty in $F = \{\rho =\sum_k
c_{k,0}P_{k,0} + \sum_k c_{k,1}P_{k,1}\}$. The face $F$ contains
$\rho_{line \,\ell=0}$ and $\rho_{line \,\ell=1}$, and the edge
joining them $\{\alpha \rho_{line \,\ell=0} +(1-\alpha )\rho_{line
\,\ell=1}\}$.

The general matrix, see (\ref{lsplit}), in $F$ is $\rho = A_{\ell=0} \oplus A_{\ell=1}\oplus 0$.
It is transformed by PT, see (\ref{ptm}), to three times
$$B=\frac13  \left( \begin{array}{ccc}  d_0 & 0 & a_1^{*} \\
                  0 & d_1 & a_0\\ a_1 & a_0^{*} & 0 \end{array} \right).$$
This is only then positive, if $a_0=a_1=0$.
That, in turn, implies, by (\ref{defma}), $c_{0,0}=c_{1,0}=c_{2,0}=\alpha$, and $c_{0,1}=c_{1,1}=c_{2,1}=1-\alpha$.
All other $\rho$ are NPT, hence entangled.

The cases of centers of triangular faces not corresponding to a
phase space line are represented by
$c_{0,0}=c_{1,0}=c_{0,1}=\frac13$, with non-vanishing $a_0$ and
$a_1$.$\Box$\\
\\
In geometric terms, this theorem is a statement about the $N-1$ dimensional faces of $\W$,
spanned by $N$ vertices with Bell type states.
Up to $N=4$ there appear at most single separable points.
For $N=5$ and $N=6$ we have to distinguish the equivalence classes.
Already for $N=5$, in the class not contained in a face $F$ treated above,
there appear new problems: Around the center, the state with coefficients
$c_{0,0}=c_{1,0}=c_{2,0}=c_{0,1}=c_{0,2}=\frac15$, there is a full four-dimensional
set of states, which are outside the kernel polytope, but PPT. This can be observed
by looking at the matrix $B$, obtained by PT of this center. It has the coefficients
$d_0 =\frac35$, $a_0=0$, $a_1 =a_2 =d_1 =d_2=\frac15$. These coefficients allow for small
variations, without destructing the positivity of $B$.

In the next application, looking into the interior of $\W$, we study
entanglement of mixtures of two orthogonal Bell type states and
$\omega$. Note the generality of this case: We use the methods
developed for $\W$, but, as it follows from our
Theorem~\ref{bellandw}, we can choose any pair of mutually
orthogonal Bell states, without reference to any special version of
$\W$. To apply our methods, we represent the state as \beq \rho =
\frac{1-(\alpha+\beta)}9 \one +\alpha P_{1,0} +\beta
P_{2,0},\qquad\textrm{with}\quad
\{1+8\alpha-8\beta\geq0,1-\alpha+8\beta\geq0\}\,. \eeq This gives,
besides $a_1 =a_2 =0$, the non-vanishing matrix elements \beqa
d_0 &=&\frac{1+2(\alpha+\beta)}3 \nonumber \\
d_1=d_2&=&\frac{1-(\alpha+\beta)}3 \nonumber \\
a_0&=&-\frac{\alpha+\beta}2 +i\frac{\sqrt{3}}2
(\beta-\alpha)\,.\nonumber \eeqa Now, $\rho$ is PPT, iff $B\geq 0$
\quad \beq \Leftrightarrow \quad 0 \leq d_1^2-|a_0|^2 =
 \frac19 \left( 1-2(\alpha+\beta)-\frac54(\alpha+\beta)^2\right)
 -\frac34(\beta-\alpha)^2\,.
\eeq
This describes, when the inequality is replaced by an equality, a nonlinear border between PPT and NPT states.
Special points on this border are:
\begin{itemize}
\item {\bf isotropic states}, $\beta=0$, border point at $\alpha=\frac14$
\item {\bf middle line}, $\alpha=\beta$, border point at $\alpha+\beta=\frac25$
\end{itemize}

Now, PPT is, for qutrits, no longer sufficient to guarantee
separability, see Refs.\cite{H96,H98,Benatti}; ``bound
entanglement'' may occur. So we have to use more specialized
methods, to study the faces not covered by the
Theorem~\ref{lowface}, and to analyze the interior of $\W$.

\subsection{Constructions of witnesses}\label{witness}

An entanglement witness, $EW_\rho$, gives a criterion, to show that
a certain state with density matrix $\rho$ is not contained in
\textrm{SEP}, the set of separable states; see Ref.~\cite{T01}. \beq
\{EW_\rho\}= \{K=K^{\dag} \,\,|\,\, \forall \sigma \in
\textrm{SEP}:\quad \Tr (\sigma K)\geq 0,\, \Tr (\rho K)< 0 \}\,.
\eeq In this paper we are mostly interested in the structure of
\textrm{SEP}. It is a convex set, and as such completely
characterized by its tangential hyperplanes. The tangents itself are
at the border of a larger set of hyperplanes which do not cut
\textrm{SEP}. So we extend the meaning of {\it witness} and define
SW, the set of {\bf structural witnesses}: \beq SW= \{K=K^{\dag}\neq
0 \,\,|\,\, \forall \sigma \in \textrm{SEP}:\quad \Tr (\sigma K)\geq
0,\ \}\,. \eeq

Similarly, we define $TW_\rho$,  the set of {\bf tangential
witnesses} for a state on the surface of \textrm{SEP} \beq TW_\rho =
\{K=K^{\dag}\neq 0 \,\,|\,\, \forall \sigma \in \textrm{SEP}:\quad
\Tr (\sigma K)\geq 0,\,\, \Tr (\rho K)= 0 \}\,. \eeq The set $SW$ is
convex and closed. It is also a linear cone: $K\in SW,\, a\geq 0,\,
\Rightarrow aK\in SW$. Therefore the bounded set $\{K\in SW, \ \Tr
K^{\dag}K \leq 1\}$ contains all the mathematical information about
$SW$; especially, that its boundary, that is $TW$, is closed. Also,
that for every $K$ the family $\{\alpha K-(1-\alpha)\omega\}$ has to
intersect the boundary at some $TW_\rho$. In geometric terms: In
each family of parallel hyperplanes, in the Euclidean space of
hemitean matrices, there are two tangential planes. Moreover,
because of convexity, closedness and boundedness of \textrm{SEP}:
For every $K$ in the boundary of $SW$, there exists at least one
$\rho \in \textrm{SEP}$, such that $K$ is a $TW_\rho$. The boundary
of \textrm{SEP} is our object of main interest.

Here we analyze $\textrm{SEP}\cap \W$, and the symmetries of $\W$
are an important tool. That a symmetry of a state is reflected in
symmetries of its witnesses seems intuitively clear, and has been
used already, Ref.~\cite{N06}. We use this correspondence of
symmetries in several details, so we formulate it explicitly:
\begin{thm}\label{witsymm}
Consider a symmetry group $\G$, implemented by unitary and/or antiunitary operators $V_g$.
Suppose that $\rho$ is $\G$-invariant, t.i. $\forall g\,\ V_g \rho V_g^{-1} =\rho$.
\begin{enumerate}
\item If $\rho$ is entangled, there exists a $\G$-invariant $EW_\rho$.
\item If $\rho$ is on the surface of \textrm{SEP}, there exists a $\G$-invariant $TW_\rho$.\label{ib}
\item The subset of $\G$-invariant elements of \textrm{SEP} is completely characterized by the
      subset of $\G$-invariant elements on the surface of SW, which is the set of $\G$-invariant $TW_\rho$.
\item All the same is true, when \textrm{SEP} is replaced by the set of \textrm{PPT}-states, entanglement by \textrm{NPT}.
\end{enumerate}
\end{thm}
\textbf{PROOF} We use the symmetrizing {\it twirl} operation, see
Ref.~\cite{VW00}, $K\mapsto \langle K \rangle_\G$. Here we use only
finite discrete groups, so the Haar-measure is just summation, and
\beq \langle K \rangle_\G = \frac1{|\G |} \sum_g  V_g K V_g^{-1},
\eeq where $|\G |$ is the number of elements in $\G$. For every
invariant $\rho$ we have
$$\Tr(K\rho )= \Tr(K\langle \rho \rangle_\G) =\Tr(\langle K\rangle_\G \,\rho).$$
If $K$ is an $EW_\rho$, then also $\langle K\rangle_\G$ is an
$EW_\rho$, proving (1). If $\rho$ is on the surface of \textrm{SEP},
there exists a $TW_\rho$, say $K$, such that $\Tr(K\rho)=0$. Then
also $\langle K\rangle_\G$ is a $TW_\rho$, proving (2). The set SW
is convex and spanned by all convex combinations of $TW_\rho$. The
same is true for the subset of invariant structural witnesses; they
are spanned by all convex combinations of invariant tangential
witnesses, on the surface. The invariant density matrices form also
a closed convex set, of lower dimension. It is completely
characterized by its tangential hyperplanes in this lower
dimensional space. These are given through the invariant $SW$'s.
This proves (3). To prove (4), one observes that the PPT states form
a closed convex set, and that the distinction between PPT and NPT is
invariant under unitary and antiunitary transformations. $\Box$\\
\\
For applications using the symmetries inside of $\W$, we
use later also the phase space reflections, implemented by local antiunitaries.
So we had to consider also this kind of group action.

As first application we consider the group of unitaries $U_{k,\ell}
=2P_{k,\ell}-1$ and their products. They are reflections,
$U_{k,\ell}^2=\one$. Our object $\W$ is pointwise invariant under
this group, and its linear span is the largest set with this
property. So, to study witnesses characterizing $\W$, we have to
consider the invariant opterator\beq
K=\sum_{k,\ell}\kappa_{k,\ell}P_{k,\ell}\, . \eeq This is sufficient
to obtain all facts about the  structure of \textrm{SEP} and PPT in
$\W$. Note: $K$ is an $EW_\rho$ for some state, iff at least one
$\kappa_{k,\ell}<0$.

Now, once more, we use the ``magic'' of Bell type states.
\begin{thm}\label{thmwitness}
The operator
\beq
K=\sum_{k,\ell}\kappa_{k,\ell}P_{k,\ell}
\eeq
is a structural witness iff $\forall \phi \in \C^3$ the operator
\beq
M_\phi =
\sum_{k,\ell}\kappa_{k,\ell}W_{k,\ell}|\phi\rangle\langle\phi|W_{k,\ell}^{-1}
\eeq
is not negative.

$K$ is moreover a $TW$ for some $\rho\in\W$, iff $\,\exists\phi$, such that $\det M_\phi =0$.
\end{thm}
\textbf{PROOF} Each separable state is a convex combination of pure
product states $|\psi ,\eta\rangle\langle\psi,\eta|$. So $K$ is a
$SW$, iff
$$\forall\psi,\eta \quad\langle \psi,\eta|K|\psi,\eta\rangle \geq 0.$$
Now we use the definitions defined in Sect.~\ref{const}
$$P_{k,\ell}=\frac13\sum_{s,t}W_{k,\ell}|s,s\rangle\langle
t,t|W_{k,\ell}^{-1}$$ and calculate
$$\langle \psi,\eta|P_{k,\ell}|\psi,\eta\rangle =$$
$$=\frac13\sum_{s,t}\langle \psi|W_{k,\ell}|s\rangle\langle\eta|s\rangle
\langle t|\eta\rangle\langle t|W_{k,\ell}^{-1}|\psi\rangle $$
$$=\frac13\langle \psi|W_{k,\ell}|\phi\rangle \langle \phi|W_{k,\ell}^{-1}|\psi\rangle,$$
giving \beq\label{kmphi} \langle \psi,\eta|K|\psi,\eta\rangle
=\frac13 \langle \psi|M_\phi|\psi\rangle\, . \eeq Here we defined
the vector $\phi\in\C^3$ as $|\phi\rangle=\sum_s
\langle\eta|s\rangle|s\rangle$, with the complex conjugated
expansion coefficients of $\eta$. What is true for all $\phi$ is
then obviously true for all $\eta$, and vice versa.

If $\det M_\phi=0$, there exists an eigenvector $\psi$ with
eigenvalue $0$. So $K$ is $TW_\rho$ for a density matrix we can
define explicitly by  $$\rho= \langle |\psi,\eta \rangle\langle
\psi,\eta| \rangle_\G\,.$$ We use here the group $\G$ with the
$U_{k,\ell}$ which we used in the first application of
Theorem~\ref{witsymm}, so $\rho\in\W$. On the other hand, if
$\exists \rho$ such that $K$ is $TW_\rho$, one may expand
$\rho=\langle |\psi,\eta \rangle\langle \psi,\eta| +\rho_{rest}$,
with $\rho_{rest}$ also being separable. Then
$$\Tr K\rho = \frac13 \langle\psi|M_\phi|\psi\rangle+\Tr K\rho_{rest}=0.$$
Each of the contributions has to vanish; so $\psi$ is an eigenvector
of $M_\phi$ with eigenvalue $0$, and $\det M_\phi =0$. $\Box$\\
\\
First application of the Theorem~\ref{thmwitness}: The well known
optimal $EW$ for a Bell type state. Consider $P_{0,0}$. We use the
symmetry of the phase space sub-group where we fix one point, e.g.
$(k,\ell )=(0,0)$, and mix all the other phase space points. An
invariant witness has to have the form $K= \gamma P_{0,0}+\beta
\one$. This gives \beq M_\phi =\gamma
|\phi\rangle\langle\phi|+3\beta \|\phi\|^2 \one\,. \eeq We have used
the representation of the unit operator on the global Hilbert space
as $\one=\sum_{k,\ell}P_{k,\ell}$. This gives then as contribution
to $M_\phi$ on $\C^3$ the operator $\sum_{k,\ell}W_{k,\ell}|\phi
\rangle\langle \phi|W_{k,\ell}^{-1}$. This operator is invariant
under the Weyl group, and its trace is $9\|\phi\|^2 $. This can only
give $3\|\phi\|^2 \one$, as the contribution to $M_\phi$. The
eigenvalues of $M_\phi$, for normed $\phi$, are $\gamma+3\beta$,
$\beta$,$\beta$. So, if $\gamma=-3\beta$, $K$ is the
\textbf{isotropic} witness  $TW_\rho$ for \beq \rho=\frac14 P_{0,0}
+\frac34 \omega\,. \eeq

In this determination of $EW$, the choice of $\phi$ was completely irrelevant.
This is connected with the high symmetry of $P_{0,0}$.

In a next application we consider fewer symmetries. We use the same
methods as in Sect.~\ref{inside}, in the equations from
(\ref{tilde}) to (\ref{lift}). Let $\G$, implemented by {\it local}
unitaries or antiunitaries $V_g\tilde{V}_g^{-1}$, be the invariance
group for $\rho$, an element of $\W$. Choosing an invariant witness
$K$, associated to the set of matrices $M_\phi$, then every $M_\phi$
is unitarily equivalent to $M_\chi$, with $\chi=V_g^{-1} \phi$. This
is seen by applying $V_g\tilde{V}_g^{-1}$ from left and its inverse
from the right onto $K$ in equation(\ref{kmphi}), and calculating
its action onto the matrix $M_\phi$.

For states and their witnesses which are located on a line in phase
space, with an eventual part proportional to $\omega$ or $\one$,
this brings essential simplification. By equivalence, we may
consider the line $\{(0,0)\ldots(2,0)\}$. All the $W_{k,0}$,
consequently the $P_{k,0}$, and of course also the unit operator,
are invariant under the group of unitaries $\sum_s
e^{i\delta(s)}|s\rangle\langle s|$. The consequence is, that each
$M_\phi$ is equivalent to $M_{|\phi|}$ with real valued non negative
vector $|\phi|$.

For general witnesses there remain the phase space translations $V_g$,
combined with $\tilde{V}_g$ as local unitary operators.
They act onto the Weyl operators by multiplication with phase factors
which cancel in the action onto $K$. The consequence for equivalences of $M_\phi$ are
the symmetries of $\det M_\phi$ under cyclic permutation $\phi_s \mapsto \phi_{s+1}$,
and the discrete phase twirling $\phi_s \mapsto w^s \phi_s$. One knows therefore, that the determinant
depends on the $\phi_s$ and their conjugates in the form of symmetric polynomials,
invariant under the discrete phase twirling. This allows permutation-symmetric sums
with contributions from $|\phi_s|^2$, from
$\phi_0^2\phi_1^{*}\phi_2^{*}$, etc. But it forbids contributions as $\phi_s^2$,
$\phi_0^2\phi_1\phi_2^{*}$ etc.

Combining these results, it is not difficult to calculate the
determinants for witnesses located on the phase space line $\ell=0$,
mixed with $\one$: Consider \beq\label{kdef} K=\lambda\frac13\one
+\sum_k \gamma_k P_{k,0}\, , \eeq related, when $\|\phi\|=1$ to the
matrices $M_\phi = \lambda\one + \sum_k\gamma_k
W_{k,0}|\phi\rangle\langle\phi| W_{k,0}^{-1}$. The matrix is written
explicitly in the appendix. The determinant works out as
\beqa\label{determinant}
\det M_\phi = \lambda^3 &+& \|\phi\|^2\left(\gamma_0+\gamma_1+\gamma_2\right)\lambda^2\\
       &+&3\left(|\phi_0|^2|\phi_1|^2+|\phi_1|^2|\phi_2|^2+|\phi_2|^2|\phi_0|^2\right)
       \left(\gamma_0\gamma_1+\gamma_1\gamma_2+\gamma_2\gamma_0\right)\lambda\nonumber\\
       &+&27|\phi_0|^2|\phi_1|^2|\phi_2|^2\gamma_0\gamma_1\gamma_2\,.\nonumber
\eeqa
The analysis follows in the next sections.

\begin{figure}
\center{\includegraphics[width=350pt,
keepaspectratio=true]{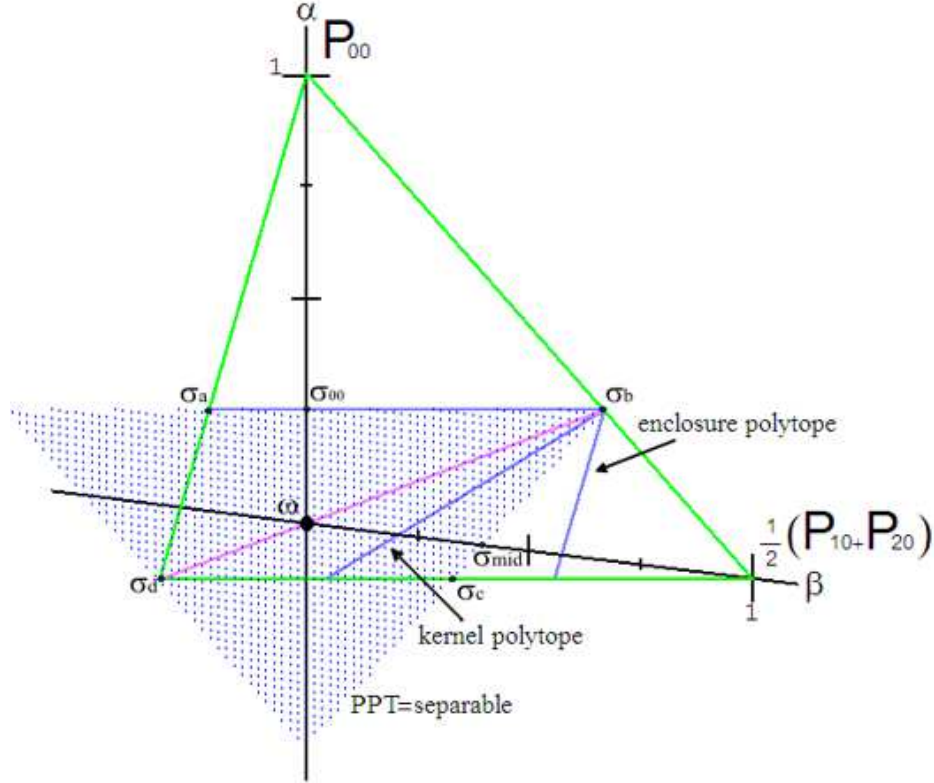}} \caption{(Color online.) Here the
space of the density matrices
$\rho=\frac{1-\alpha-\beta}{9}\;\omega+\alpha P_{00}+\frac{\beta}{2}
(P_{10}+P_{20})$ is shown. The green triangle shows the positivity
condition. The blue lines are the inner fence (kernel polytope) and
the outer fence (enclosure polytope) for the boundary of
separability. The dotted area shows the values of $\alpha$ and
$\beta$ which are positive under PT. Here PPT equals
separability.}\label{bild2}
\end{figure}

\subsection{Some details in the structure,
analytical}\label{detailsa}

The strategy for the exploration of the structure of \textrm{SEP} is
to find the set of tangential witnesses as follows: Analyse the
operators $K=\sum_{k,\ell}\kappa_{k,\ell}P_{k,\ell}$ by way of
studying the set of matrices $M_\phi$ associated to each single $K$.
If these matrices are positive for all $\phi$, then $K$ is a $SW$.
If there is a $\phi$, such that $\det M_\phi=0$, then $K$ is a $TW$.
If one has ``enough'' $TW_\rho$s, one can determine the $\{\rho\}$
in the boundary of \textrm{SEP}. Consequently, they obey $\Tr K\rho
=0$. How many of these witnesses are ``enough'', depends on the
symmetry of the subset of states one is studying. High symmetry
restricts and simplifies the study.

We study the subset of states with components located on a {\bf
phase space line}, mixed with $\omega$. By equivalence, it is
sufficient to study one special line. We choose that with $\ell=0$.
The states are
$\rho=\alpha\sum_k\kappa_{k,0}P_{k,0}+(1-\alpha)\omega$ with $\sum_k
\kappa_{k,0}=1$. Each of these states is invariant under vertical
shear and horizontal reflection. This symmetry group implies that we
may restrict the study of witnesses to \beq\label{linewit}
K=\lambda\frac13 \one + \sum_k\gamma_kP_{k,0}\,, \eeq as at the end
of Sect.~\ref{witness}. Note that the parameter $\lambda$ cannot be
negative for witnesses, since $\Tr K\rho_{line\,\ell\neq
0}=\lambda/3$, but $\Tr K\sigma$ should be non-negative for
separable states. And we know already, that all the $\rho_{line}$
are separable, Theorem~\ref{twelve}.

Especially simple are those operators, where $\lambda=0$:
$K$ is a $TW$, iff all $\gamma_k\geq 0$.
Because, if all the factors $\gamma_k$ are non-zero, then each $M_\phi$ is a sum of positive operators.
But, if one of them is negative, then, with $\phi=\sum_s \frac1{\sqrt{3}}|s\rangle$,
the matrix $M_\phi$ has a negative eigenvalue.
Such a $K$ with non-negative $\gamma_k$ is no $EW$, but tangential to the face of $\W$ spanned by the $P_{k,\ell\neq0}$.

In the study of the operators with $\lambda >0$, it is enough to consider
$\lambda=1$, because the witnesses form a cone, all parameters may be scaled.
The investigation, whether $M_\phi\geq 0$, is now done by investigation of the characteristic
polynomial $\Xi_c(\mu)=\det(M_\phi-\mu)$. Since $M_\phi$ is hermitian, this polynomial,
 $\Xi_c(\mu)=\prod_j (\mu_j-\mu)$,
has only real zeroes $\mu_j$.
We have to demand that they are not negative, and this is the case if and only if
\begin{itemize}
\item the second derivative of $\Xi_c$ at $\mu=0$ is not negative,
\item The first derivative there is not positive,
\item $\Xi_c(\mu=0)$ is not negative.
\end{itemize}
These are conditions for $SW$. To get $TW$, we need one eigenvalue equal to zero,
and strengthen the last condition to
\begin{itemize}
\item $\Xi_c(\mu=0)=0$.
\end{itemize}
The characteristic polynomial is given by replacing $\lambda$ with $1-\mu$ in the equation (\ref{determinant}).
We use real valued $\phi$ and the abbreviations
\beqa
\begin{array}{cccccc}
A&:=& \gamma_0\gamma_1+\gamma_1\gamma_2+\gamma_2\gamma_0, &\quad\quad
B&:=& \gamma_0\gamma_1\gamma_2\\
f_A &:= &3(\phi_0^2\phi_1^2+\phi_1^2\phi_2^2+\phi_2^2\phi_0^2), &\quad\quad
f_B &:=& 27\phi_0^2\phi_1^2\phi_2^2 \label{efb}
\end{array}\,.
\eeqa The conditions for a tangential witness are given by
calculating the derivative $\Xi_c$, using $\|\phi\|=1$,
\begin{itemize}
\item $3+\sum_k\gamma_k \geq 0$,
\item $3+2\sum_k\gamma_k +\min_\phi (A\cdot f_A) \geq 0$,
\item $1+\sum_k\gamma_k +\min_\phi(A\cdot f_A+B\cdot f_B)=0$.
\end{itemize}
The minima over normalized wavefunctions $\phi$ are evaluated in the appendix:
\beqa
\min_\phi (A\cdot f_A) &=& \min\{0,A\}, \label{mina}\\
\min_\phi(A\cdot f_A+B\cdot f_B)&=& \min\{0, \frac34 A,\, A+B\}\,.
\eeqa Because of (\ref{mina}), the first of the conditions as stated
above for witnesses is weaker then the second one. And the
parameters for $TW$s have to fulfill only one inequality and one
equation: \beqa 3+2\gamma_0+4\gamma +\min\{0,A\} &\geq 0
\label{inequ}
\\
1+\gamma_0+2\gamma + \min\{0, \frac34 A,\, A+B\} &=0\,; \label{cond}
\eeqa where we use now the parameters \beq \gamma
=\frac12(\gamma_1+\gamma_2),\quad\quad \delta
=\frac12(\gamma_1-\gamma_2)\,. \eeq Using them we get \beq
A=2\gamma_0\gamma+\gamma^2-\delta^2,\quad\quad
A+B=2\gamma_0\gamma+\gamma^2+\gamma_0\gamma^2\,. \eeq

 In the search for $TW_\rho$ for $\rho$ symmetric under vertical
 reflection, i.e. $\kappa_{1,0}=\kappa_{2,0}$,
one can restrict the search to operators $K$ which have the same symmetry, that is, they have $\delta=0$.
We explore the set of witnesses starting from the isotropic witness, with $\lambda=1$,
$\gamma_0=-1$, $\gamma=\delta=0$.
First we list all those $TW$s one gets, then we indicate the ``proof''.

In the set of results we find four distinguished regions for the
parameters:
\begin{description}
\item[a)] $\lambda=1$,\quad\quad $\gamma\geq 0$,\quad\quad \quad\quad $\gamma_0=-1$;\label{a}
\item[b)] $\lambda=1$,\quad\quad $0\geq \gamma \geq -\frac23$,\quad  $\gamma_0=-1-2\gamma\leq \frac13$;
\item[c)] $\lambda=1$,\quad\quad  $\gamma = -\frac23$,\quad\quad\quad $\gamma_0\geq \frac13$;
\item[d)] $\lambda=0$,\quad\quad  $\gamma \geq 0$,\quad\quad\quad\quad  $\gamma_0=1-\gamma>0$.
\end{description}
In the parameter region {\bf a)} one has $A=\gamma^2-2\gamma\geq-1$,
so the l.h.s. of Eq.~(\ref{inequ}) is positive; and
$\min\{0,\frac34A,A+B\}=-2\gamma$, so Eq.~(\ref{cond}) is true. At
one end of the region, i.e. in the limit $\gamma \to \infty$, one
may rescale the parameters and observe that they approach
$\lambda=\gamma_0=0$, $\gamma>0$, one end of region {\bf d)}. At the
other end of region {\bf a)}, which is also the beginning of region
{\bf b)}, the common witness at the edge of these regions is the
isotropic witness with $\gamma=0$ and $\gamma_0=-1$, corresponding
to the hyperplane $B_{0,0}$, defined in Eq.~(\ref{isoplane1}) and in
Eq.~(\ref{isoplane2}). In the parameter region {\bf b)}, succeeding
{\bf a)}, neither $A$ nor $A+B$ is negative, since $\gamma\leq 0$,
as long as  $\gamma\geq -\frac23$, with $\gamma_0$ related to
$\gamma$ by (\ref{cond}). In the succeeding region {\bf c)} one has
$A\leq 0$, and $\frac34 A\leq A+B$. At the end of this region
rescaling leads here, in the limit $\gamma_0\rightarrow \infty$, to
$\lambda=\gamma=0$, $\gamma_0>0$, this is the other end of region
{\bf d)}. The round trip is finished.

For regions {\bf a)}, {\bf b)} and {\bf c)}, we get $EW$s, except for $\gamma=\gamma_0=-\frac13$
in region {\bf b)}. There we get the $TW_\rho$ for all the $\rho$ in the
triangular face of $\W$ with the $P_{k,0}$ at the vertices.

The linearity of the relations implies, that in each regional set of
witnesses each $K$ is a $TW_\rho$ for (at least) one common $\rho$.
For each $K$ at and endpoint of a region, there exists a linear face
of \textrm{SEP}, for which $K$ is tangential~\footnote{The relation
between the boundary of a convex set like \textrm{SEP} and its
tangential hyperplanes generalizes the relation between a concave or
convex function and its Legendre transform. This relation is well
known in physics, especially concerning the thermodynamic functions.
And there this duality between linear regions and singular vertex
points is known from the phase transitions.}.

The vertex points of \textrm{SEP}, corresponding to the linear
regions of the witness-parameters, are:
\begin{description}
\item[a)] $\sigma_a=\omega+\frac29P_{0,0}-\frac19P_{1,0}-\frac19P_{2,0}$;\label{rhoa}
\item[b)] $\sigma_b=\rho_{line \, \ell =0}$;
\item[c)] $\sigma_c=\frac34(\omega-\frac19P_{0,0}+\frac29P_{1,0}+\frac29P_{2,0})$;
\item[d)] $\sigma_d=\frac32(\omega -\rho_{line \,\ell =0})=\frac12(\rho_{line\, \ell =1}+\rho_{line \,\ell =2})$.
\end{description}
The hyperplane $B_{0,0}$, corresponding to the isotropic witness, contains
four $\rho_{line}$, associated with the four phase-space-lines through the phase-space point $P_{0,0}$.
One of them is $\sigma_b$; the other three have $\sigma_a$ in their middle:
$$\sigma_a =\frac13\left(\frac13(P_{0,0}+P_{0,1}+P_{0,2})+\frac13(P_{0,0}+P_{1,1}+P_{2,2})+
\frac13(P_{0,0}+P_{2,1}+P_{1,2})\right).$$ The part of the boundary
of \textrm{SEP} between $\sigma_b$ and $\sigma_c$ is outside the
kernel polytope and inside the enclosure polytope. It crosses the
ray from $\omega$ to $\frac12(P_{1,0}+P_{2,0})$ at
$\sigma_{mid}=\frac15(3\omega+P_{1,0}+P_{2,0})$.

\begin{figure}
\center{\includegraphics[width=350pt,
keepaspectratio=true]{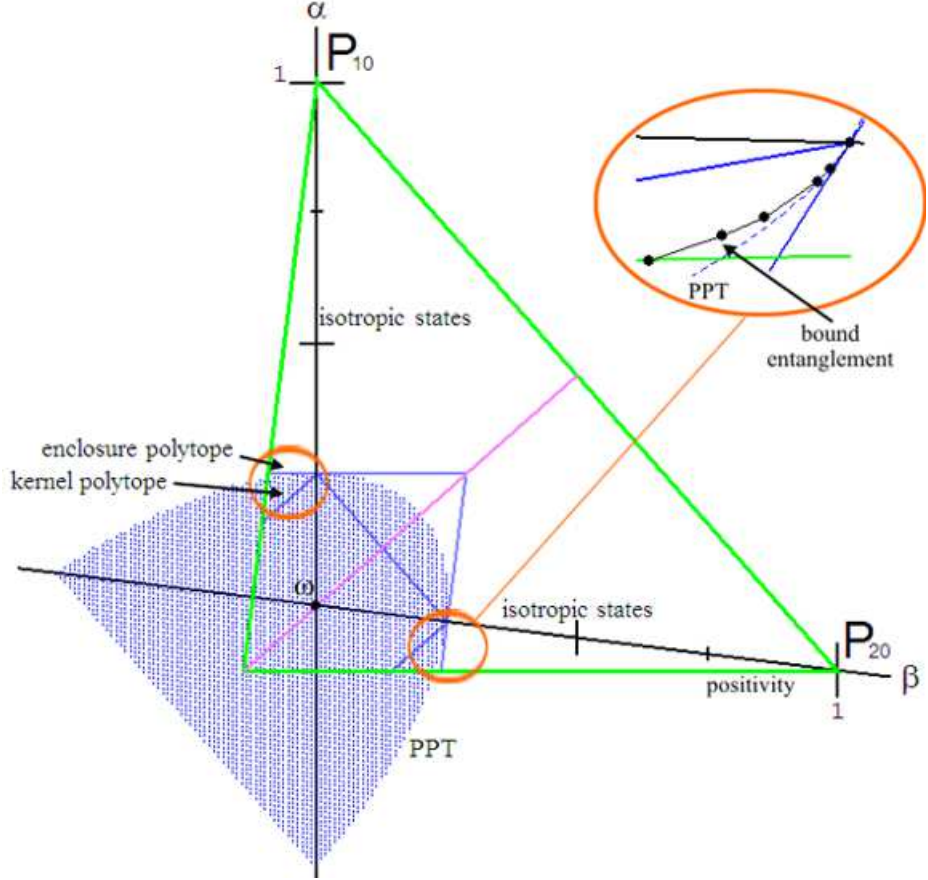}} \caption{(Color online.) Here the
space of the density matrices
$\rho=\frac{1-\alpha-\beta}{9}\;\omega+\alpha P_{10}+\beta P_{20}$
is shown. The green triangle shows the positivity condition. The
axes joining $\omega$ and $P_{10}$ or $P_{20}$ represent the
isotropic states. The blue lines are the inner fence (kernel
polytope) and the outer fence (enclosure polytope) for the boundary
of separability. The dotted area shows the region of $\alpha$ and
$\beta$ where $\rho$ is PPT. For both parameters $\alpha, \beta\geq
0$ PPT gives the boundary between separable and entangled states.
For $\alpha$ or $\beta$ negative we find bound entanglement which
can be seen in the enlarged picture. The dashed curve shows the
border of PPT and for the points the witnesses were explicitly
numerically calculated, see Sect.~\ref{detailsn}.
}\label{bild}
\end{figure}

\subsection{Some details in the structure, numerical}\label{detailsn}

Our method of numerical analysis is a variation of the strategy we
use in the previous section. There we calculate first the whole set
of tangential witnesses $K$, then we find the $\rho$ on the border
of \textrm{SEP} via the condition $Tr(\rho K)=0$. Here we use this
condition from the beginning and reduce in this way the set of
parameters which have to be varied. We find that explicitly in the
following way.

Let us here consider again the density matrix
$\rho=\frac{1-\alpha-\beta}{9}\omega+\alpha P_{1,0}+\beta P_{2,0}$
where the state space is visualized in Fig.~\ref{bild}.  Via
Theorem~\ref{thmwitness} the operator $K=\mathbbm{1}+a P_{00}+b
P_{10}+c P_{20}$ is a structural witness iff the matrix
$M_\phi=\mathbbm{1}+a W_{0,0}|\phi\rangle\langle\phi|W_{0,0}+b
W_{1,0}|\phi\rangle\langle\phi|W_{1,0}+c
W_{2,0}|\phi\rangle\langle\phi|W_{2,0}$ is non negative $\forall
\phi$. We are moreover interested in the tangential witness,
therefore we search for $Tr(\rho K)=0$ which leads to
$a=\frac{3+b(1+8 \alpha-\beta)+c(1-\alpha+8\beta)}{\alpha+\beta-1}$.
Consequently, we have to search for the border where for a given
$\alpha$ and the variation over $b, c$ and $\phi$  one eigenvalue of
$M_\phi$ changes from negative to positive (the two others are
positive). The found minimal $\beta$ characterizes then the border
state for which a separable state changes to an entangled state, and
$K$ is the tangential entanglement witness. We did not include any
further symmetry constraints into the calculation and found that the
analytical symmetry results as described in the previous section and
in the appendix are confirmed, e.g. we have only to vary over real
vectors $\phi$. In the region where no bound entanglement was found,
the optimized parameter $\beta$ agreed with the PPT calculated
$\beta$ numerically up to $10^{-8}$. The largest difference between
the PPT boundary and the separability boundary is of the order of
$10^{-2}$ and decreases to zero for $\alpha$ or $\beta$ approaching
$0$, the isotropic state, see Fig.~\ref{bild}.

\section{Summary and conclusions}

We consider the state space of two qutrits where we restrict
ourselves to locally maximal mixed states. Whereas for qubits every
locally maximally mixed state can be diagonalized by the magic Bell
states and therefore embedded into a magical tetrahedron, this is
not true for qutrits. However, we show that a kind of analogue to
the magic tetrahedron can be defined for qutrits: the magical
simplex $\W$.

Starting from a certain maximally entangled pure state, a Bell type
state, we obtain by applying only on Alice side the Weyl operators
nine other orthogonal Bell type states. The Weyl operators are used
to describe the discrete classical phase space. This discrete
classical phase space representing the algebraic relations of the
Weyl operators enables us to describe the local transformations of
the state space of interest and are very useful for several proofs
in this paper. The mixtures of all Bell type states form the simplex
$\W$ which is then the main object of our investigations.

We show explicitly how to construct a version of $\W$. A certain
version is fixed by defining $3$ Bell type states. The simplex $\W$
can be embedded in a $9$--dimensional Euclidean space equipped with
a Hilbert-Schmidt norm and an inner product.

Transformations of $\W$ onto itself can be considered as
transformations of the discrete classical phase space. Thus the
symmetries and equivalences can be studied by this means.

Then we investigate the question of the geometry of separability. We
start with constructing two polytopes, an inner (kernel polytope)
and an outer (enclosure polytope) fence for separability. They
define linear entanglement witnesses but are in general not optimal.
The outer fence, the closure polytope, has the same geometric
symmetry as $\W$.

Hereupon we explicitly study two representative cases. We consider
the state space of all density matrices which are mixtures of the
total mixed state and two Bell states. We apply the partial positive
transformation on one subsystem (PPT) which detects entanglement.
The obtained witness is no longer a linear one. We show how
entanglement witnesses can be constructed and apply it to the
density matrices under consideration. We find after optimizing the
entanglement witness by analytical and independently by numerical
methods that there is indeed bound entanglement for negative
mixtures of one of the Bell states. The result is also visualized in
Fig.~\ref{bild}.

The second case we study is the state space of density matrices
which are mixtures of the total mixed state, one Bell state and an
equal mixture of two other Bell states. We find that it has only
linear entanglement witnesses and that no bound entanglement can be
found, visualized in Fig.~\ref{bild2}.

Summarizing, we could give a full geometric structure of the subset
of bipartite qutrits under investigation. We think that this will
help to find a good characterization of the whole state space and to
investigate measures for entanglement for higher dimensional
systems.

\section*{Acknowledgements}
B.C. Hiesmayr wants to acknowledge the EU-project EURIDICE
HPRN-CT-2002-00311.

\section{Appendix}

\subsubsection{A physical model}

Each party has a system consisting of a ring shaped molecule. In this ring there are
three symmetric located possible places for a single itinerant particle. Locating this particle at any of
these places corresponds to our three basis vectors $|s\rangle$.
In the entangled state, described by the vector $\Omega_{k,\ell}$, the index $\ell$
denotes the angular correlations between Alice's and Bob's particles.
Concerning measuring of locations, for $\ell =0$ they are to be measured at the places at the same angles.
For the two other cases, Alice's particle is rotated relative to Bob's.
So the number $\ell$ is the quantum number for the observable $s_{Bob}-s_{Alice}$;
and the index $k$ is the quantum number for the total angular momentum (component orthogonal to the rings).
Again  ``$-1$''$=2$, due to the finiteness of the system.
These two operators commute, while their individual contributions from one party do not;
comparable to the commuting of $x_B - x_A$ with $p_B + p_A$.
The set of their eigenvalue pairs ${(k,\ell )}$ is the {\bf discrete classical phase space}.

\subsubsection{Examples of maximally mixed states which do not fit into $\W$}

An example for the observation (\ref{undec}) in Sect.~\ref{place}:
Define $\rho = \frac{1}{3}|\Psi\rangle\langle\Psi | +
\frac{2}{3}|\Phi\rangle\langle\Phi |$, with $|\Psi\rangle =
|0\rangle \otimes  |0\rangle $, \quad and $|\Phi\rangle =
\frac{1}{\sqrt{2}}(|1\rangle \otimes |1\rangle + |2\rangle \otimes
|2\rangle )$. This density matrix is in a unique way diagonalized,
-- with non Bell states.

As an example for the observation (\ref{unw}) we consider $\rho =
\sum c_\alpha |\Psi_\alpha \rangle\langle\Psi_\alpha |$, with three
different orthonomal normalized Bell vectors $\Psi_\alpha$ and three
different expansion factors $c_\alpha$. Since such an expansion is
just the expansion into projectors onto eigenvectors, it is unique.
So, {\it if} $\rho$ can be embedded into $\W$ (or a unitary
equivalent), the expansion must be an expansion into the
$P_{k,\ell}$ (or into a unitarily equivalent set).

Now we give an example of three Bell type projectors which cannot
together be embedded into $\W$: Take two of the projectors as
$P_{0,0}$ and $P_{1,0}$, the third one as $|\Phi\rangle\langle\Phi
|$, with $|\Phi\rangle = \third \sum_{s}
w^{2s}\frac{1}{3}(2|s\rangle + 2|s-1\rangle -|s+1\rangle)\otimes
|s\rangle$. This is a Bell vector, orthogonal to $\Omega_{0,0}$ and
$\Omega_{1,0}$. But it is not orthogonal to $\Omega_{2,0}$, so
$|\Phi\rangle\langle\Phi |$ cannot be any $P_{k,\ell}$. Now try a
transformation and embed the first two projectors as
$P_{0,0}=P'_{k,\ell}$ and $P_{1,0}=P'_{k+m,\ell +n}$ into a unitary
equivalent version of $\W$. Consider the mapping between these
projectors by Weyl operators. As the following equation shows, they
are fixed up to a phase \beq P_{1,0}= P'_{k+m,\ell +n} = (U\otimes
\one)P'_{k,\ell}(U\otimes \one)^{\dag} = (U\otimes
\one)P_{0,0}(U^{\dag}\otimes \one)\,. \eeq Taking the matrix
elements with $|s\rangle \otimes |t\rangle$, identifying $U$ with
$W'_{m,n}$ and comparing with the relation between $P_{0,0}$ and
$P_{1,0}$ gives the equations \beq \langle s| W'_{m,n} |t\rangle =
\langle s| U |t\rangle = e^{i\eta}\langle s| W_{1,0} |t\rangle =
e^{i\eta} w^s \delta_{s,t}\,. \eeq And this implies, by applying
$W'_{m,n}$ once more, that $P'_{k+2m,\ell +2n}=P_{2,0}$, and the
third Bell state does not fit into the equivalent version of $\W$
either.

An example for the observation (\ref{ninebell}):
Take the three Bell states $\Omega_{k,2}$ out of $\W$ and replace them by
$$\Phi_k = \third \sum_s w^{ks} \alpha_s |s-\ell \rangle \otimes |s\rangle\,.$$
They are orthogonal, span the same subspace as the deleted $\Omega_{k,2}$, but define other states,
unless the phase factors $\alpha_s$ are chosen in a very special way.
Together with the remaining $\Omega_{k,0}$ and $\Omega_{k,1}$ they form
a complete set of orthogonal Bell vectors, but nothing equivalent to $\W$.

\subsubsection{Matrices representing the Weyl operators}

The basis vectors are \beq |0\rangle =
\left(\begin{array}{c}1\\0\\0\end{array}\right) , \quad |1\rangle =
\left(\begin{array}{c}0\\1\\0\end{array}\right) , \quad |2\rangle =
\left(\begin{array}{c}0\\0\\1\end{array}\right)\,. \eeq The Weyl
operators $W_{k,\ell}$, arranged according to the appearance of the
indices in the phase space are \beqa
\begin{array}{ccccccc}k=2\quad &\left( \begin{array}{ccc}1&0&0\\0&w^{*}&0\\0&0&w \end{array}\right) , \quad
                     &\left( \begin{array}{ccc}0&1&0\\0&0&w^{*}\\w&0&0 \end{array}\right) , \quad
                     &\left( \begin{array}{ccc}0&0&1\\w^{*}&0&0\\0&w&0 \end{array}\right) , \quad \\
                     \\
                   k=1\quad &\left( \begin{array}{ccc}1&0&0\\0&w&0\\0&0&w^{*} \end{array}\right) , \quad
                     &\left( \begin{array}{ccc}0&1&0\\0&0&w\\w^{*}&0&0 \end{array}\right) , \quad
                     &\left( \begin{array}{ccc}0&0&1\\w&0&0\\0&w^{*}&0\end{array}\right) , \quad \\
                     \\
                   k=0\quad &\left( \begin{array}{ccc}1&0&0\\0&1&0\\0&0&1 \end{array}\right) , \quad
                     &\left( \begin{array}{ccc}0&1&0\\0&0&1\\1&0&0 \end{array}\right) , \quad
                     &\left( \begin{array}{ccc}0&0&1\\1&0&0\\0&1&0 \end{array}\right) , \quad \\
                     \\
                    \quad \ell =  &  0 & 1 & 2
                   \end{array}\,.
\eeqa

Complex conjugation interchanges the lines $k=2$ and $k=1$.

The transformation producing unitary operators are represented as
\beq U_R = \frac{1}{\sqrt{3}} \left(
\begin{array}{ccc}1&1&1\\1&w^{*}&w\\1&w&w^{*} \end{array} \right),
\qquad  U_V =            \left(
\begin{array}{ccc}1&0&0\\0&1&0\\0&0&w^{*} \end{array} \right)\,. \eeq

\subsubsection{The partial transposition}
The basis elements are the product vectors $|s-\ell ,s\rangle$. They are arranged in groups of three,
according to $\ell$. Inside each group the ordering is according to $s$. The index-pairs $(s-\ell ,s)$
denoting the rows are written on the left side.
Partial transposition induces a new splitting of the Hilbert space in subspaces. They are characterized by
$m$, when we write the index-pairs now as $(m -s ,s)$. We mark the different $m$ by different typefaces;
$\bf{a}$ for $m=0$, $\rm{a}$ for $m=1$, $a$ for $m=2$.
(But the numbers are the same, independent of the typeface!)
\beqa
                 \begin{array}{c}
                                  (0,0)\\
                                  (1,1)\\
                                  (2,2)\\
                                  \\(2,0)\\
                                  (0,1)\\
                                  (1,2)\\
                                  \\(1,0)\\
                                  (2,1)\\
                                  (0,2)\\

 \end{array} \quad
 \left(  \begin{array}{ccccccccccc}

                                  \bf{d_0}&\rm{a_0^{*}}&a_0& &0&0&0& &0&0&0\\
                                  \rm{a_0}&d_0&\bf{a_0^{*}}& &0&0&0& &0&0&0\\
                                  a_0^{*}&\bf{a_0}&\rm{d_0}& &0&0&0& &0&0&0\\
                                  \\
                                  0&0&0& &d_1&\bf{a_1^{*}}&\rm{a_1}& &0&0&0\\
                                  0&0&0& &\bf{a_1}&\rm{d_1}&a_1^{*}& &0&0&0\\
                                  0&0&0& &\rm{a_1^{*}}&a_1&\bf{d_1}& &0&0&0\\
                                  \\
                                  0&0&0& &0&0&0& &\rm{d_2}&a_2^{*}&\bf{a_2}\\
                                  0&0&0& &0&0&0& &a_2&\bf{d_2}&\rm{a_2^{*}}\\
                                  0&0&0& &0&0&0& &\bf{a_2^{*}}&\rm{a_2}&d_2\\
 \end{array}\right)
 \qquad\bf{\mapsto}
 \eeqa
 \beqa
 \begin{array}{c}  \\
                                  (0,0)\\
                                  (1,1)\\
                                  (2,2)\\
                                  \\(2,0)\\
                                  (0,1)\\
                                  (1,2)\\
                                  \\(1,0)\\
                                  (2,1)\\
                                  (0,2)\\

 \end{array} \quad
 \left(  \begin{array}{ccccccccccc}

                                  \bf{d_0}&0&0& &0&0&\bf{a_2}& &0&\bf{a_1^{*}}&0\\
                                  0&d_0&0& &a_2&0&0& &0&0&a_1^{*}\\
                                  0&0&\rm{d_0}& &0&\rm{a_2}&0& &\rm{a_1^{*}}&0&0\\
                                  \\
                                  0&a_2^{*}&0& &d_1&0&0& &0&0&a_0\\
                                  0&0&\rm{a_2^{*}}& &0&\rm{d_1}&0& &\rm{a_0}&0&0\\
                                  \bf{a_2^{*}}&0&0& &0&0&\bf{d_1}& &0&\bf{a_0}&0\\
                                  \\
                                  0&0&\rm{a_1}& &0&\rm{a_0^{*}}&0& &\rm{d_2}&0&0\\
                                  \bf{a_1}&0&0& &0&0&\bf{a_0^{*}}& &0&\bf{d_2}&0\\
                                  0&a_1&0& &a_0^{*}&0&0& &0&0&d_2\\
 \end{array}\right)\,.
\eeqa

\subsubsection{The matrix $M_\phi$ for witnesses on a line, mixed with the unit}
We use real valued $\phi_s$. This is possible because of the
invariances as described at the end of the Sect.~\ref{witness}: \beq
\left(
\begin{array}{ccccc}\lambda+\phi_0^2(\gamma_0+\gamma_1+\gamma_2)&
&\phi_0\phi_1(\gamma_0+w^{*}\gamma_1+w\gamma_2)&&
                                      \phi_0\phi_2(\gamma_0+w\gamma_1+w^{*}\gamma_2)\\
                       \phi_1\phi_0(\gamma_0+w\gamma_1+w^{*}\gamma_2)&&\lambda+\phi_1^2(\gamma_0+\gamma_1+\gamma_2)&&
                                      \phi_1\phi_2(\gamma_0+w^{*}\gamma_1+w\gamma_2)\\
                       \phi_2\phi_0(\gamma_0+w^{*}\gamma_1+w\gamma_2)&&\phi_2\phi_1(\gamma_0+w\gamma_1+w^{*}\gamma_2)&&
                                        \lambda+\phi_2^2(\gamma_0+\gamma_1+\gamma_2)
\end{array}\right)\,.
\eeq

\subsubsection{The minima for the functions of $\phi$, used in Sect.~\ref{detailsa}}

We use real valued, normalized wavefunctions $\phi$ and describe them with the two parameters
\beqa
z&=&\phi_0^2 \quad \in [0,1],\label{zet}\\
x&=&\frac12(\phi_1^2-\phi_2^2)\quad \in [-\frac{1-z}2,
+\frac{1-z}2]\,.\label{ex} \eeqa The minima of the functions of
$\phi$ can be evaluated as minima of functions of $z$ and $x$ in the
triangular region determined in (\ref{zet}) and (\ref{ex}). The
functions defined in the Sect.~\ref{detailsa} in Eq.(\ref{efb}) are
\beqa
f_A&=& \frac34(1+2z-3z^2)-3x^2,\\
f_B&=&27z\left(\frac{(1-z)^2}4-x^2\right)\,. \eeqa Because of the
terms quadratic in $x$, the extrema are attained either at the
boundary of the triangle, or at the line $x=0$. One finds the
minimum of $f_A$ at the triangle-vertices, equal to $0$. The maximum
is attained in the center, at $z=\frac13$, equal to $1$. Now one has
to respect the sign of $A$: \beq \min_{x,z}(A\cdot
f_A)=\min\{0,A\}\,. \eeq In the combinations of $f_A$ with $f_B$
also the maximum of $f_A$ at the border of the triangle comes into
consideration. It is found at the center of a border line and it is
equal to $\frac34$.

The non negative function $f_B$ is zero along the whole boundary and
has its maximum, equal to $1$, at the center.
The minimum of $Af_A+Bf_B$ is zero, attained at a vertex, if both $A$ and $B$ are non-negative.
It is attained at the center, and equal to $A+B$, if both $A$ and $B$ are non-positive.
For  different signs of $A$ and $B$ one has to analyse $Af_A+Bf_B$ along the line $x=0$. Consider
$$f_A+Cf_B = \frac34(1-z)\left(1+3z+9Cz(1-z)\right)$$
as functions of $z$, depending on the parameter $C=B/A$.
For every $C$ they have fixed values at $z=0$ -- there they are $\frac34$ -- and at $z=1$ -- where
they are $0$. For every $C$ the first derivative at $z=\frac13$ is zero. There is either a local minimum
 -- if $C<-\frac13$ --,
a local maximum -- if $C>-\frac13$ --, or a saddle point.
Now we consider the function with special values of the parameter $C$:
For $C=-1$, it is zero at $z=1/3$, which is a local minimum.
For $C=-1/4$, the local maximum at $z=1/3$ is equal to the maximal value $\frac34$ at the border, at $z=0$.
Inside the region $0<z<1$
the set of functions is pointwise monotone increasing in the parameter $C$.
So for $C$ in between the special values, the minimum is zero, the maximum is $\frac34$, both attained at the border.
For $C\leq -1$, the minimum, $1+C$, is attained at the center, the maximum, $\frac34$, at the border.
For $C\geq -\frac14$, the minimum, zero, is attained at a vertex, the maximum, $1+C$, at the center.

The result can be stated in a unified way as the equation \beq
\min_{x,z}(A\cdot f_A+B\cdot f_B)=\min\{0,\frac34 A, A+B\}\,. \eeq
It means, that the search for the minimum over all $\phi$ can be
reduced to considering only three different vectors: \beq \left(
\begin{array}{c} 1\\0\\0 \end{array} \right), \quad\quad
\frac1{\sqrt2}\left( \begin{array}{c} 1\\1\\0 \end{array} \right),
\quad\quad \frac1{\sqrt3}\left( \begin{array}{c} 1\\1\\1 \end{array}
\right)\,. \quad\quad \eeq

\end{document}